 \documentclass{emulateapj}

\usepackage{apjfonts}

\slugcomment{to appear in the Astrophysical Journal, Letters}

\shorttitle{Supersoft X-ray Light Curve of RS Oph}
\shortauthors{Hachisu et al.}

\begin{document}


\title{Supersoft X-ray Light Curve of RS Ophiuchi (2006)}


\author{Izumi Hachisu}
\affil{Department of Earth Science and Astronomy,
College of Arts and Sciences,
University of Tokyo, Komaba 3-8-1, Meguro-ku, Tokyo 153-8902, Japan}
\email{hachisu@chianti.c.u-tokyo.ac.jp}

\author{Mariko Kato}
\affil{Department of Astronomy, Keio University, 
Hiyoshi 4-1-1, Kouhoku-ku, Yokohama 223-8521, Japan}
\email{mariko@educ.cc.keio.ac.jp}

\and

\author{Gerardo Juan Manuel Luna}
\affil{Instituto de Astronomia, Geof\'{\i}sica e Ci\^encias
Atmosf\'ericas, Universidade de S\~ao Paulo,
Rua do Mat\~ao 1226, Cid. Universitaria, 05508-900, S\~ao Paulo, Brasil}
\email{gjmluna@astro.iag.usp.br}




\begin{abstract}
One of the candidates for Type Ia supernova progenitors, the
recurrent nova RS Ophiuchi underwent the sixth recorded outburst
in February 2006, and for the first time a complete light curve of
supersoft X-ray has been obtained.  It shows the much earlier emergence
and longer duration of a supersoft X-ray phase than expected before.
These characteristics can be naturally understood when a significant
amount of helium layer piles up beneath the hydrogen burning zone
during the outburst, suggesting that the white dwarf (WD) is effectively
growing up in mass.  We have estimated the WD mass in RS Oph
to be $1.35 \pm 0.01~ M_\odot$ and the growth rate of the WD mass
to be at an average rate of about $1 \times 10^{-7} M_\odot$ yr$^{-1}$.
The white dwarf will probably reach the critical mass
for Type Ia explosion if the present accretion continues
further for a few to several times $10^{5}$ years. 
\end{abstract}





\section{Introduction}
Recurrent novae are binary star systems in which mass is transferred
onto a white dwarf (WD) primary from a main-sequence or a red giant
secondary.  The eruption is well-modeled as a thermonuclear runaway
(hydrogen shell flash), which occurs when a certain amount of mass
($\Delta M_{\rm ig}$) is accumulated on the surface of the white dwarf
\citep[e.g.,][]{pri95}.  From their short recurrence periods
(from a ten to several tens of years) and very rapid optical declines,
it is believed that their white dwarfs are very massive
and close to the Chandrasekhar mass \citep[e.g.,][]{hac01kb}.
If the WD mass increases
after every outburst, it will soon explode as a Type Ia supernova
\citep[e.g.,][]{nom82, hkn99, hknu99, hac01kb}.  It is, therefore,
crucially important to know how close the WD mass is
to the Chandrasekhar mass and how much mass is left on the white dwarf
after one cycle of nova outburst.

The WD envelope rapidly expands and blows winds
\citep[e.g.,][]{kat94h} as a result of hydrogen-flash,
and its photospheric radius reaches a maximum and then gradually shrinks.
Since the total luminosity is almost constant during the outburst,
the photospheric temperature increases in time.  We easily understand
this from Stefan-Boltzmann's equation,
$L_{\rm ph} = 4 \pi R_{\rm ph} \sigma T_{\rm ph}^4$,
where $L_{\rm ph}$, $R_{\rm ph}$, and $T_{\rm ph}$ are the photospheric
luminosity, radius, and temperature, respectively.
The main emitting region moves from optical to ultraviolet, and then
finally to supersoft X-ray, corresponding to from $T_{\rm ph} \sim 10^4$K
through $T_{\rm ph} \sim 10^6$K.
The photosphere drastically shrinks at the end of the wind phase and
a supersoft X-ray phase starts \citep[e.g.,][]{kat99}.
We are able to constrain the WD mass and its
growth rate if turn-on/turnoff of a supersoft X-ray phase
are detected, because they indicate the durations of
wind mass loss (how much mass is ejected)
and hydrogen shell burning without wind mass loss (how much mass
is left).

\begin{figure}
\epsscale{1.15}
\plotone{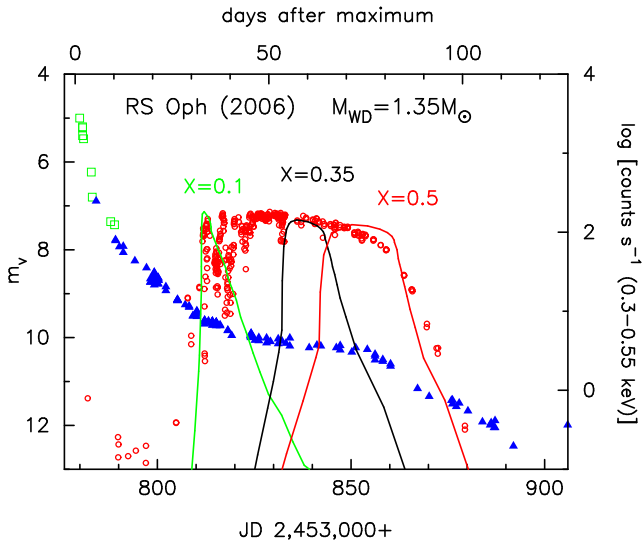}
\caption{
Observed X-ray count rates and theoretical X-ray flux
for an adiabatic boundary model.
{\it Open circles}: Supersoft X-ray count rates of
the {\it Swift} XRT observations, which are extracted
in the $0.3-0.55$~keV energy range.
{\it Solid lines}: Model X-ray fluxes
for $M_{\rm WD}= 1.35~M_\odot$ are plotted against time
(days after the optical peak, i.e., JD~2453778.79; 2006 February 12.29 UT).
Hydrogen content ($X=0.1$, $X=0.35$, and $X=0.5$)
is attached to each curve.
Model X-ray flux is calculated as a simple blackbody
with the response function of the {\it Swift} XRT
between $0.3$ and $0.55$~keV, i.e., a band of $22.5-41.3$~\AA~. 
We shift vertically the X-ray flux curve to fit the peak of
the count rate, because we intend to approximately explain the duration
of the supersoft X-ray peak but not the detailed behavior.
So, no interstellar absorption is assumed in our model X-ray light curves.
{\it Filled triangles and open squares}: A visual light curve of
the 2006 outburst \citep[taken from ][]{hac06b}, i.e.,
$V$ ({\it open squares}) and $y$ ({\it filled triangles}) magnitudes.
\label{no_helium_heat}
}
\end{figure}

The sixth recorded outburst of RS Oph was discovered on 2006 February 12
by a Japanese amateur astronomer H. Narumi \citep{nar06}
when it was shining at 4th magnitude.  Figure \ref{no_helium_heat}
shows the optical development of the outburst \citep[taken from][]{hac06b}.
The visual light decayed rapidly during the first week
and then the decay gradually slowed down (``early decline phase'').
It remained at about 10th magnitude from 40 to 80 days after the
optical maximum, i.e., for about 40 days (``mid-plateau phase'').
The final decline started about 80 days after the optical maximum
(``final decline phase'').
It eventually decayed to 12th magnitude (``post-outburst minimum''),
which is a magnitude darker than that in usual quiescent phase.

A very bright supersoft X-ray phase of RS Oph was extensively
observed by the {\it Swift} XRT \citep[e.g.,][]{bod06a}.
\citet{osb06a} reported the emergence of highly variable
soft X-ray flux between 30 and 40 days after the optical maximum.
Then the supersoft X-ray flux was stabilized at about 40 days
and reached a maximum at about 50 days, followed by
a linear decline between 60 and 80 days from $\sim 200$ to
$\sim 100$ counts~s$^{-1}$ for $0.2-10$ keV \citep{osb06b}.
It rapidly declined at about 90 days.
Thus the duration of supersoft X-ray phase is about 60 days
\citep{osb06b}.

In this Letter, we calculate theoretical light curve
models based on the optically thick wind theory and 
reproduce both the supersoft X-ray and optical light curves.
The main difference from our previous models \citep{hac06ka} is
to include the effect of heat exchange between the hydrogen burning
layer and the helium layer built up beneath the hydrogen burning zone.

     In \S \ref{x-ray_data_rs_oph}, we present a complete light curve
of supersoft X-ray for the \object{RS Oph} 2006 outburst.
The numerical model of our theoretical light curves and their fittings
with the observation are presented in \S \ref{model_xray_rs_oph}.
Discussion and conclusions follow in \S \ref{discussion}.

\section{X-ray DATA} \label{x-ray_data_rs_oph}

We analyze {\it Swift} XRT observations available in the
HEASARC\footnote{http://heasarc.nasa.gov/}
database following the standard procedures.
Observations were taken in PC (photon counting) and WT
(windowed Timing) modes depending on the arrival count rate.
Source photons were extracted from a circular region
of $40\arcsec$ in PC mode and 40 $\times$15 pixels in WT mode.
No background subtraction was performed as it
represented less than 1\% of the emission during the
analyzed phases of the outburst and therefore its
effect was negligible. In Figure \ref{no_helium_heat}, we summarize an
average value of the supersoft X-ray count rates in
every 2000s bin. A narrow energy band of 0.3-0.55 keV
was adopted in order to avoid possible contamination
by a tail of hard X-ray photons coming from high
temperature plasma at the shock \citep[e.g.,][]{bod06a}.

\citet{osb06a} reported an oscillatory behavior in the {\it Swift} XRT
count rate during 30 and 31 days after the optical maximum,
corresponding to the period in which the supersoft X-ray count rate
first quickly rose (Fig. \ref{no_helium_heat}). 
The count rate began to rapidly decline at about 80 days, which
is coincident with the final decline of the optical light curve
as pointed out by \citet{hac06b}.

The optical mid-plateau phase of \object{RS Oph} is first clearly
identified by the $y$ light curve in the 2006 outburst \citep{hac06b}.
Such mid-plateau phases are also
observed in the other recurrent novae, \object{U Sco} and \object{CI Aql},
which is interpreted as a disk irradiated by the central hot white dwarf
\citep[e.g.,][]{hkkm00, hac03k}.  The irradiated disk is bright when
the central white dwarf is as luminous as that for hydrogen shell-burning
phase, and it becomes dark when the shell-burning ends.
Therefore the fact that the end of a supersoft X-ray phase is
coincident with the end of a mid-plateau phase clearly indicates that
this corresponds to the termination of hydrogen shell-burning
on the white dwarf at about 80 days after the optical maximum.

\begin{figure}
\epsscale{1.15}
\plotone{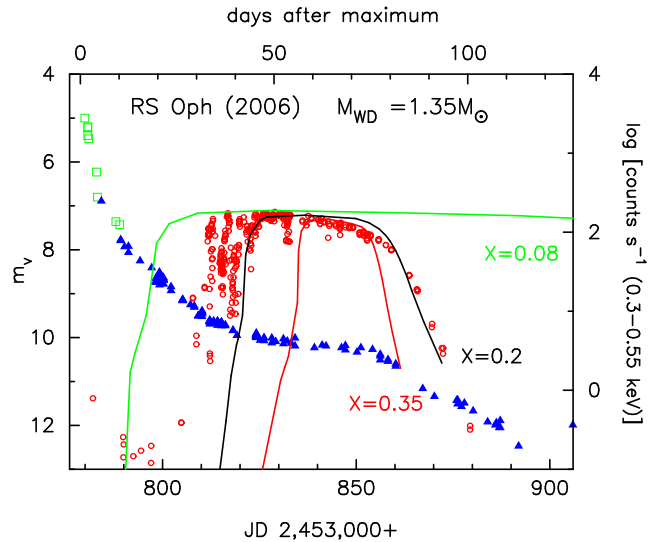}
\caption{
Same as those in Fig.\ref{no_helium_heat}
but for models in which heat exchange
between the hydrogen burning zone and helium layer is taken into account.
\label{yes_helium_heat}
}
\end{figure}

\section{Model of supersoft X-ray phase} \label{model_xray_rs_oph}


We calculate nova light curves based on the optically
thick wind theory \citep{kat94h}.
Our theoretical model is consisting of a white dwarf,
a disk around the white dwarf, and a red giant companion.
Irradiation effects of each component are included \citep{hac01kb}.

In the previous paper \citep{hac06ka}, we predicted the duration of
a supersoft X-ray phase of the \object{RS Oph} 2006 outburst.
Figure \ref{no_helium_heat} shows our calculated supersoft X-ray
fluxes by using the same method 
as in the previous paper, i.e.,
we assumed an adiabatic condition at the bottom
of the hydrogen burning layer \citep{kat94h},
so heat did not flow inward but only outward.
We cannot reproduce the early emergence
(day $30-40$) and long duration (60 days)
of the supersoft X-ray phase at the same time.

Here we introduce a new treatment by which we include the effect of
heat exchange between the hydrogen burning zone and helium layer.
After the optical maximum, convection descends in time.  So
no more processed helium is carried upward but it accumulates
underneath the burning zone. The accumulated helium layer keeps
a large amount of thermal energy because the temperature of burning zone
is as high as $\sim 10^8$~K.
The temperature of burning zone gradually decreases
in the later phase of the outburst.
Then heat flux from the hot helium layer becomes important
in the luminosity.  This heat flux from the thermal reservoir makes
the lifetime of a supersoft X-ray phase much longer.
Figure \ref{yes_helium_heat} represents such new light curves for
$X=0.08$, $0.20$, and $0.35$.  It is very clear that both the earlier
emergence and longer duration of a supersoft X-ray phase are realized
at the same time.

\begin{figure}
\epsscale{1.15}
\plotone{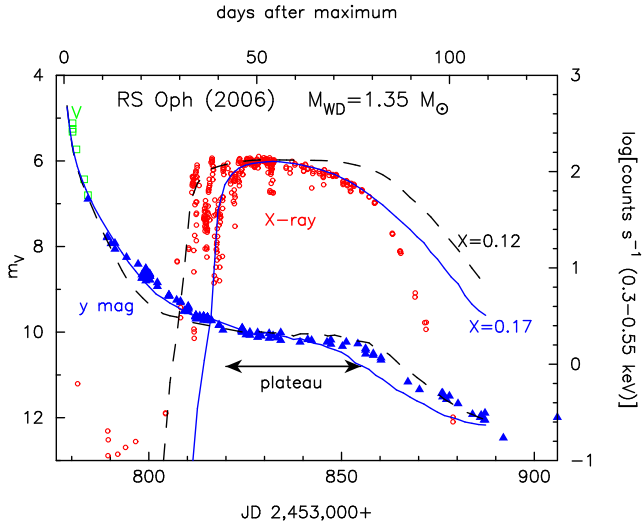}
\caption{
Same as those in Fig. \ref{yes_helium_heat}, but together
with the visual light curves.
Hydrogen content in the envelope is $X=0.12$ ({\it dashed lines})
and $X=0.17$ ({\it solid lines}).
These two models predict visual light curves as shown in the figure.
Here we assume a $0.7~M_\odot$ red giant with a radius of $35~R_\odot$
for the companion, an irradiated disk with a radius of $47 ~R_\odot$
around the white dwarf, a binary orbital period of 455.72 days, and
a binary inclination angle of $i=33^\circ$ \citep{hac06b}.
\label{rsoph2006_ssxs}
}
\end{figure}

Our numerical model should reproduce not only the supersoft X-ray light
curve but also the visual light curve.
Figure \ref{rsoph2006_ssxs} compares the
observational visual light curve with our numerical ones.
In the early decline phase, free-free emission from the optically thin
ejecta 
dominates the visual light \citep{hac06b, hac06kb}.
The irradiation effects of the disk and the companion red giant
dominate the optical light in the mid-plateau phase, whereas
the WD photosphere does not contribute
because it shrinks to $0.005-0.01 ~R_\odot$,
much smaller than the disk.  The final decline started
at about 80 days, corresponding to the decay of hydrogen-burning.

In the model with $M_{\rm WD}=1.35 ~M_\odot$ and $X=0.17$,
optically thick winds stopped at $t_{\rm wind}= 40$ days
after the optical peak.
At this epoch, hydrogen of $0.16 \times 10^{-7} M_\odot$ still exists
and can supply more 14 days luminosity by nuclear burning.
The heat stored in the helium layer,
$\sim 5 \times 10^{44}$ ergs, can also supply
a total luminosity of $2.8 \times 10^{38}$ ergs s$^{-1}$ for another
20 days to the supersoft X-ray phase.  Therefore these energies can
maintain the supersoft X-ray phase for a total of 34 days.  As a
result, steady hydrogen shell-burning on the white dwarf ends roughly at
$t_{\rm steady}= 75$ days.

Another critically important value, the mass accretion rate, can be
estimated as follows:  In our model with $M_{\rm WD}= 1.35 ~M_\odot$
and $X=0.17$ ($X=0.12$), the envelope mass at the optical peak
is $\Delta M_{\rm
ig} \sim 4 \times 10^{-6} M_\odot$.  This implies that the average
mass accretion rate onto the white dwarf is $\dot M_{\rm acc} \sim 2
\times 10^{-7} M_\odot$ yr$^{-1}$ during the quiescent phase between
1985 and 2006.  Here we neglect a dredge-up of core materials
mainly because carbon and oxygen was not enriched in the ejecta.
Among the accreted matter, the wind carries away about 70\% (50\%), i.e.,
$\Delta M_{\rm wind} \sim 2.8 \times 10^{-6} M_\odot$
($\Delta M_{\rm wind} \sim 2 \times 10^{-6} M_\odot$);
this is much larger than the X-ray
observational indication of $\sim 1 \times 10^{-7} M_\odot$ by
\citet{sok06}, but roughly consistent with the infrared
observational indication of $\sim 3 \times 10^{-6} M_\odot$
\citep{das06, lan07}.  Note that the wind duration is
$t_{\rm wind}= 40$ days (32 days).
The residual, $\sim 1.2 \times 10^{-6} M_\odot$
($\sim 2 \times 10^{-6} M_\odot$),
accumulates on the white dwarf. So the white dwarf is growing at an
average rate of $\dot M_{\rm He} \sim 0.6 \times 10^{-7} M_\odot$
yr$^{-1}$  ($\dot M_{\rm He} \sim 1 \times 10^{-7} M_\odot$
yr$^{-1}$) during $1985-2006$.




\begin{figure}
\epsscale{1.15}
\plotone{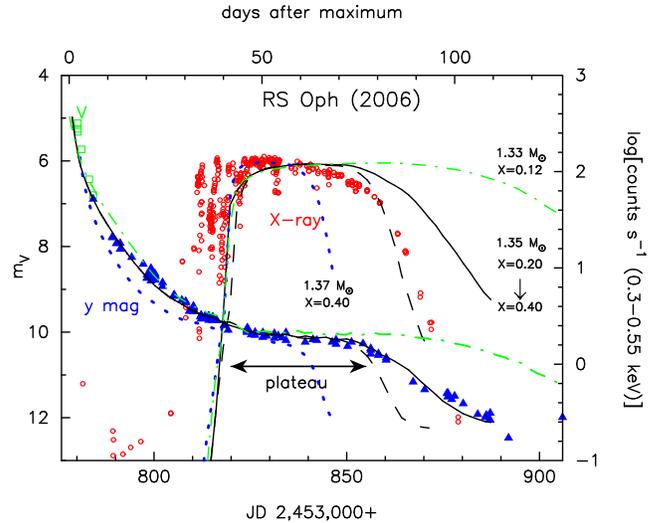}
\caption{
Same as those in Fig. \ref{rsoph2006_ssxs} but for different
WD masses and chemical compositions:
$M_{\rm WD}= 1.37~M_\odot$, $X=0.40$ ({\it dotted line});
$M_{\rm WD}= 1.33~M_\odot$, $X=0.12$ ({\it dash-dotted line});
and a non-uniform chemical composition model ({\it solid line}), where
we increase the hydrogen content from $X_{\rm nuc}=0.20$
to $X_{\rm nuc}=0.40$ in the nuclear burning zone only 
during the plateau phase for $M_{\rm WD}= 1.35~M_\odot$.
A uniform chemical composition model of $M_{\rm WD}= 1.35~M_\odot$
with $X=0.20$ ({\it dashed line}) is added for comparison.
\label{rsoph2006_ssxs_mass}
}
\end{figure}

\section{Discussion and conclusions}
\label{discussion}
In the present work, we reproduce the duration of a supersoft X-ray phase
by introducing heat flux from the helium layer
underneath the hydrogen burning zone.
In our previous works 
\citep{kat94h, kat99, hac01kb, hac06ka, hac06kb, hac06b},
we had included heat flux from the helium layer
only during the cooling phase, for simplicity.
When the WD mass is not so massive 
($M_{\rm WD} \lesssim 1.0 ~M_\odot$) 
and the hydrogen content is not so small ($X \gtrsim 0.3$),
this effect can be neglected because only a small part of nuclear
energy is absorbed into the helium layer.  However, 
we found that this effect cannot be neglected 
when the WD mass is so massive as the Chandrasekhar mass
($\sim 1.3 ~M_\odot$) and the hydrogen content is very small
($X \lesssim 0.3$), as seen in Figures \ref{no_helium_heat} and
\ref{yes_helium_heat}.

We have assumed that the chemical composition is uniform throughout
the envelope for simplicity.
As mentioned earlier, a gradient of the hydrogen content 
is reasonable if convection descended in time during the rising
phase of the nova outburst.  
  Although the hydrogen content
of RS Oph ejecta has not been estimated in detail, a very low
hydrogen content $X \sim 0.1$ was reported in the late phase of the U Sco
1979 outburst \citep{bar81}.  An interesting trend was also found
in the latest 1999 outburst of \object{U Sco}:
\citet{iij02} reported
a rather high value of $X \sim 0.6$ at 16 hours after the optical
maximum and \citet{anu00} obtained a value of $X \sim 0.4$
at $11-12$ days after the optical maximum.
The hydrogen content, $X$, may not be uniform
but gradually increase outward.

We assume $X=0.20$ but, in the plateau phase,
we increase the hydrogen content from $X_{\rm nuc}=0.20$
to $X_{\rm nuc}=0.40$ at the hydrogen burning zone, where
$X_{\rm nuc}$ is the hydrogen content at the nuclear burning zone.
As shown in Figure \ref{rsoph2006_ssxs_mass},
more hydrogen content extends the supersoft X-ray phase.

In the previous work \citep{hac06b}, we have estimated the WD 
mass to be $1.35 \pm 0.01 ~M_\odot$ from the optical light curve fitting.
Here we have estimated the WD mass and obtained
the same results, that is, $1.35 \pm 0.01 ~M_\odot$.
A more massive white dwarf ($1.37~M_\sun$)
can be rejected because the optical light
curve decays too fast in the early decline phase and the duration of
the supersoft X-ray phase is too short
(see Fig. \ref{rsoph2006_ssxs_mass}).  A less massive one
($1.33~M_\sun$)
can also be excluded because we must assume a very low hydrogen
content to reproduce the early emergence of
supersoft X-ray but its supersoft X-ray phase lasts
too long and the visual decline is too slow
as can be easily seen from Figure \ref{rsoph2006_ssxs_mass}.

The supersoft X-ray flux from our model has been calculated
from a blackbody photosphere.
Although the supersoft X-ray flux may not be a simple blackbody,
our flux is a reasonable indication of the supersoft X-ray phase,
because we intend to approximately explain the duration
of the supersoft X-ray peak (not the detailed behavior).

A few groups resolved a size of near infrared emission for the RS Oph
2006 outburst \citep{mon06, lan07, che07}.  \citet{mon06}
reported that a size of $\sim 3$ mas is consistent with the binary size
if the distance is as short as 0.6 kpc \citep{hac01kb}
but it is much larger than the binary size when the distance
is as long as 1.6 kpc \citep{hje86}.  Recently, \citet{hac06b}
revised their value and proposed a range of $d= 1.3 - 1.7$ kpc.
\citet{lan07} obtained that the angular diameter of infrared emission
increased once to $\sim 4$ mas at about day 20 and then decreased to
$\sim 2$ mas at about day 100.  Our optically thick wind model
suggests that the radius of near infrared ($2.2 \mu$) photosphere
is about 1 AU, estimated from equation (11) of \citet{wri75}, near
the optical maximum when free-free emission dominates infrared continuum.
The radius quickly decreases because the wind mass-loss rate soon
decreases and free-free emission becomes optically thin.  This maximum
radius of infrared free-free photosphere is smaller than the binary orbit
($a \sim 1.5$ AU at the distance of 1.6 kpc), so that a size of
$\sim 2-4$ mas infrared emission may originate from circumbinary matter
as discussed by \citet{mon06} and \citet{lan07}.

Thus the supersoft X-ray light curve has led us to know
various physical parameters of the white dwarf.  In this work,
the optically thick wind is essentially important
to understand the nova optical light curve and its supersoft X-ray
duration because, without winds, the nova duration is too
long to be compatible with the observation \citep{kat94h}.
Moreover, the wind duration (together with $X$)
determines the amount of ejected mass and
processed helium mass eventually left on the white dwarf.  Now we
can conclude that the WD mass, $1.35 \pm 0.01~M_\odot$,
is now increasing.
In fact the WD mass is currently increasing at 
$\dot M_{\rm He} \sim (0.5-1) \times 10^{-7} M_\odot$ yr$^{-1}$.

We may predict the future of the white dwarf.  After it has come through
many recurrent nova outbursts, the mass of the helium layer reaches
a critical mass and a helium shell flash occurs.
Its strength is weak \citep{kat99h}.
Therefore, only a small part of the helium layer
will be blown off in the wind, and virtually all of
the helium layer will be burnt into carbon-oxygen and accumulates
in the white dwarf \citep{kat99h}.
The WD mass can grow though many recurrent nova
outbursts and helium shell flashes.  If the companion will
supply matter by the same rate as the present one for another hundred
thousand years, we expect an Type Ia supernova explosion
of the white dwarf.

\acknowledgments

We acknowledge Milvia Capalbi and Lorella Angelini for their help
with {\it Swift} data reduction.
We are also grateful to the anonymous referee
for useful comments that improved the manuscript.
This research has been supported in part by Grants-in-Aid for
Scientific Research (16540211, 16540219)
of the Japan Society for the Promotion of Science.
G.J.M.L. acknowledges CNPq for his graduate
fellowship (Process 141805/2003-0).


\begin{thebibliography}{}

\bibitem[Anupama \& Dewangan (2000)]{anu00}
Anupama, G. C., \& Dewangan, G. C. 2000, \aj, 119, 1359

\bibitem[Barlow et al.(1981)]{bar81}
Barlow, M. J. et al. 1981, \mnras, 195, 61


\bibitem[Bode et al.(2006a)]{bod06a}
Bode, M. F., et al. 2006a, \apj, 652, 629




\bibitem[Chesneau et al. (2007)]{che07}
Chesneau, O. et al. 2007, \aap, in press (astro-ph/0611602)


\bibitem[Das et al. (2006)]{das06}
Das, R., Banerjee, D. P. K., \& Ashok, N. M. 2006, \apj,  653, L141






\bibitem[Hachisu \& Kato (2001)]{hac01kb}
Hachisu, I., \& Kato, M. 2001, \apj, 558, 323

\bibitem[Hachisu \& Kato (2003)]{hac03k}
Hachisu, I., \& Kato, M. 2003, \apj, 588, 1003



\bibitem[Hachisu \& Kato (2006a)]{hac06ka}
Hachisu, I., \& Kato, M. 2006a, \apj, 642, L53

\bibitem[Hachisu \& Kato (2006b)]{hac06kb}
Hachisu, I., \& Kato, M. 2006b, \apjs, 167, 59

\bibitem[Hachisu et al. (2000)]{hkkm00}
Hachisu, I., Kato, M., Kato, T., \& Matsumoto, K. 2000,
\apjl, 528, L97 

\bibitem[Hachisu et al. (2006)]{hac06b}
Hachisu, I. et al. 2006, \apj, 651, L141


\bibitem[Hachisu et al. (1999a)Hachisu, Kato, \& Nomoto]{hkn99}
Hachisu, I., Kato, M., \& Nomoto, K. 1999a, \apj, 522, 487 

\bibitem[Hachisu et al. (1999b)]{hknu99}
Hachisu, I., Kato, M., Nomoto, K., \& Umeda, H. 1999b, \apj, 519, 314

%

%

\bibitem[Hjellming et al. (1986)]{hje86}
Hjellming, R. M., van Gorkom, J. H., Seaquist, E. R., Taylor, A. R.,
Padin, S., Davis, R. J., \& Bode, M. F. 1986, \apjl, 305, L71


\bibitem[Iijima (2002)]{iij02}
Iijima, T. 2002, \aap, 387, 1013






\bibitem[Kato (1999)]{kat99}
Kato, M. 1999, \pasj, 51, 525

\bibitem[Kato \& Hachisu (1994)]{kat94h}
Kato, M., \& Hachisu, I., 1994, \apj, 437, 802

\bibitem[Kato \& Hachisu (1999)]{kat99h}
Kato, M. \& Hachisu, I. 1999, 513, L41



\bibitem[Lane et al. (2007)]{lan07}
Lane, B. F. et al. 2007, \apj, in press (astro-ph/0612099)


\bibitem[Monnier et al. (2006)]{mon06}
Monnier, J. D. et al. 2006, \apj, 647, L127



\bibitem[Narumi et al.(2006)]{nar06}
Narumi, H., Hirosawa, K., Kanai, K., Renz, W., Pereira, A., 
Nakano, S., Nakamura, Y., \& Pojmanski, G. 2006, \iaucirc, 8671


\bibitem[Nomoto (1982)]{nom82}
Nomoto, K. 1982, \apj, 253, 798






\bibitem[Osborne et al. (2006a)]{osb06a}
Osborne, J. et al. 2006a, The Astronomer's Telegram, 770, 1

\bibitem[Osborne et al. (2006b)]{osb06b}
Osborne, J. et al. 2006b, The Astronomer's Telegram, 838, 1



\bibitem[Prialnik \& Kovetz (1995)]{pri95}
Prialnik, D., \& Kovetz, A. 1995, \apj, 445, 789







\bibitem[Sokoloski et al. (2006)]{sok06}
Sokoloski, J. L., Luna, G. J. M., Mukai, K., \& Kenyon, S. J. 2006,
\nat, 442, 276








\bibitem[Wright \& Barlow (1975)]{wri75}
Wright, A. E.; Barlow, M. J. 1975, \mnras, 170, 41
%
\end{thebibliography}
\end{document}